\documentclass[12pt,fleqn]{article}
\usepackage{graphicx}
\usepackage{ifthen}
\newtheorem{law}{Law}
\newtheorem{claim}{Claim}
\newtheorem{definition}{Definition}
\title{Classes of service under perfect competition and technological change: 
a model for the dynamics of the Internet?}
\author{Daniel Lehmann\thanks{
School of Engineering and Computer Science, Hebrew University, 
Jerusalem 91904, Israel.
e-mail: lehmann@cs.huji.ac.il}
}

\begin{document}
\parindent0.0cm
\maketitle

\newboolean{color}
\setboolean{color}{false}

\begin{abstract}
Certain services may be provided in a continuous, one-dimensional, 
ordered range of different qualities 
and a customer requiring a service of quality $q$ 
can only be offered a quality superior or equal to $q$.
Only a discrete set of different qualities will be offered, 
and a service provider will provide the same service (of fixed quality $b$) 
to all customers requesting qualities of service inferior or equal to $b$.
Assuming all services (of quality $b$) are priced identically,
a monopolist will choose the qualities of service and the prices 
that maximize profit but, under perfect competition, 
a service provider will choose the (inferior) quality of service 
that can be priced at the lowest price. 
Assuming significant economies of scale,
two fundamentally different regimes are possible: 
either a number of different classes of service are offered (DC regime), 
or a unique class of service offers 
an unbounded quality of service (UC regime).
The DC regime appears in one of two sub-regimes: one, BDC, 
in which a finite number of classes is offered, 
the qualities of service offered are bounded 
and requests for high-quality services are not met,
or UDC in which an infinite number of classes of service are offered 
and every request is met.
The types of the demand curve and of the economies of scale, 
and not the pace of technological change, 
determine the regime and the class boundaries.
The price structure in the DC regime obeys very general laws.
\end{abstract}
\section{Introduction}
\label{sec:intro}
\subsection{Background}
\label{subsec:back}
Consider a delivery service. 
The quality of the service given may be measured
by the delay with which the object is delivered.
A service guaranteeing same day delivery is of higher quality
than one that guarantees only next day delivery, or delivery within 
three-business days.
Assuming that the service can be provided on a continuous range of qualities,
it is typically unrealistic to assume that a service provider could provide
all those different services to the customers requiring them or bill them
differentially for the qualities requested. In many situations, the service
provider will have to decide on certain discrete qualities, i.e., classes
of service, and will provide only services of those specific qualities.

The price paid by the different customers depends only on the quality 
of the service they receive, not on the (inferior) quality they requested
and all customers getting the same service pay the same price.
We assume that every customer has sharp requirements concerning the
quality of service he requests: he will under no circumstances accept
a quality that is inferior to his requirements, even if this is much cheaper,
and he will, as long as the price is agreeable, use the service of the
least quality that is superior or equal to his request.
This assumption is not usual, and not in line with~\cite{GibbensMasonStein:99}
for example. Nevertheless, it is very reasonable in connection with Internet
services: there is quite a sharp boundary between delays, jitters and latencies
that allow for the streaming of video content and those that do not.

We assume significant economies of scale to the service provider:
the cost of providing $w$ services grows less than linearly with $w$,
but we also assume that services of different qualities do not aggregate
to generate economies of scale: the cost of providing two classes of service
is the sum of the costs of those two different services.
This is obviously a severe assumption. 
In practice, a company offering a few different 
qualities of service will benefit from economies of scale across those
different services: administrative services for example will be shared.
But if one considers only the cost of moving packets over the Internet
and one assumes, in no way a necessary assumption but a definite possibility
(see~\cite{Odlyzko:PMPEC99}), that separate sub-networks are affected
to the different qualities of service, then the assumption will be
satisfied.
In the sequel we may therefore assume that a firm provides only a single
quality of service: different firms may provide different qualities of service.
In deciding what quality or qualities of service to provide, 
a firm has to avoid two pitfalls:
offering a service of poor quality may cater to a share of the market
that is too small to be profitable,
offering a service of high quality may involve costs that are too
high to attract enough customers.
A monopoly will set a quality of service and a price that maximize its
revenues, but,
in a competitive environment, a customer will buy from the provider offering
the lowest price for a service whose quality is superior or equal to
the quality requested. This will, in general, result in a lower quality
of service, a lower price and a higher activity.

Will more than one class of service be proposed?
How many? What will the price structure of those classes be?
The relation between the traffics in those different classes?
Between the revenues gathered in giving those services of different
qualities?

Those questions may be asked in many different situations, but they are
particularly relevant in connection to the Internet.
The Internet delivers packets of information from a source to a destination.
The IPv4 protocol, the current protocol for the Internet,
reserves three bits in each packet for specifying the quality of service 
desired, but does not use those bits and treats all packets equally.
End users rarely pay per packet and usually pay a flat rate.
A number of companies have prepared products for QoS (quality of service), 
i.e., for controlling an Internet-like network in which packets are treated
differentially, but the need for such products is not yet proven.
A survey of the different proposals for QoS may be found in~\cite{FergHust:98}.
Since the Internet is a loose organization that is the product of cooperation 
between very diverse bodies, billing for such services would also be a major
problem.
Many networking experts have therefore claimed that the current 
{\em fat dumb pipe}
model is best and argued that it will prevail due to the rapid decline
in the cost of equipment. 
A discussion of the different
predictions about the evolution of the Internet may be found
in~\cite{Odlyzko:PMPEC99}, where a specific proposal, Paris Metro Pricing
(PMP) is advocated.
Much interesting information about the economics of the Internet is found
at~\cite{Varian:website}.
An important aspect of the economy of the Internet is that the
prices of the networking equipment are dropping very rapidly. One wonders
about the consequences of this rapid decline on the price structure.
\subsection{Main results}
\label{subsec:mainresults}
Answers to the questions above are obtained, assuming a one-dimensional 
ordered continuum of qualities of service, 
significant economies of scale and a competitive environment.
Under some reasonable and quite general assumptions about the demand curve
and the cost function, 
and assuming that a change in price has a similar effect on the demand
for all qualities of service, it is shown that one obtains 
one of two situations. 
\begin{itemize}
\item (UC regime) If the demand for services of high-quality is strong
and the economies of scale are substantive, then a service provider may
satisfy requests for service of arbitrary quality: there will be only one class
of service catering for everybody's needs; this is the {\em fat dumb pipe}
model.
\item (DC regime) In other cases, a number of classes of service 
will be proposed and priced according to the quality of the service provided. 
Very high quality services may not be provided at all.
\end{itemize}

The DC regime just described may appear in one of two sub-regimes:
\begin{itemize}
\item (BDC regime) A finite number of classes of service are offered
and services of very high quality are not offered at all.
\item (UDC regime) An infinite number of classes and all qualities of service
are offered.
\end{itemize}
Notice that, for example, it cannot be the case that a finite number
larger than one of classes are offered and that the highest quality
of service caters for unbounded qualities.
Significant economies of scale and high demand for high-quality services 
imply a UC regime, independently of the way prices influence demand.
When not in a UC regime, high sensitivity of demand to price implies
a BDC regime whereas low sensitivity implies an UDC regime.
In a BDC regime, when the basic price of the equipment drops, 
new classes of service will
be offered to the high-end customers that could not be profitably catered
for previously.
In the UC regime, a slowing of the decline 
in the price of equipment, does not cause the appearance of multiple classes 
of service. In a DC regime,
a change in the price of equipment cannot cause a transition to
a UC regime.

A decline in equipment prices, in a DC regime, does not change
the boundaries between the different classes of service, 
but new classes catering to the high-end of the market become available
when prices are low enough to make them profitable.
The prices drop for all classes, but they drop dramatically for the
newly created classes. The ratio of the prices between a class of service
and the class just below it decreases and tends to some number typically
between $1.5$ and $3$, depending on the exact shape of the distribution
of demand over different qualities and the size of the economies of scale,
when prices approach zero.
Traffic increases when equipment prices decrease.
The direction in which revenues change depends on the circumstances,
but, typically, revenues go up and then down when equipment prices decrease.
The traffics in neighboring classes of service tend to a fixed ratio
when prices approach zero.

\section{Previous work}
\label{sec:previous}
This work took its initial inspiration in the model proposed by A. Odlyzko
in the Appendix to~\cite{Odlyzko:PMPEC99}.
In~\cite{Odlyzko:PMPEC99,FishOdly:98}, the authors study the case of two
different types of customers each type requesting a specific quality of
service. They show that, in certain cases, two classes of service will
be proposed.
The present work assumes a continuous distribution of types of customers.
The model of~\cite{GibbensMasonStein:99} is more detailed than the one
presented here and addresses slightly different questions, but its conclusions
are supported by the present work.

The situation described in this paper may seem closely related to
the well-studied phenomenon of non-linear pricing~\cite{Wilson:Nonlinear}.
A second look seems to indicate major differences:
\begin{itemize}
\item here the pricing is linear: a customer given many services
pays for each of those separately and, in the end, pays the sum of the prices
of the services received,
\item the cost of providing services is assumed to
exhibit significant economies of scale, whereas costs are assumed 
to be additively separable among customers and qualities
in most works on non-linear pricing,
\item the present work deals with a competitive environment whereas many
works on non-linear pricing assume a monopoly situation. 
\end{itemize}
\section{Model and assumptions}
\label{sec:model}
The results to be presented are very general, i.e., 
they hold under very mild hypotheses concerning the form of the cost function 
and the demand curve, and this is the main reason for our interest in them.
Nevertheless, describing exactly the weakest hypotheses needed for
the results to hold would lead to a paper loaded with mathematical
details distracting the reader from the essentials.
Therefore, the conditions under which those general results 
hold will not be described precisely 
and those results will be only justified by hand waving.
Precise results will deal with two specific parametrized families of functions
for cost and demand, described in~(\ref{eq:paramf}) 
and~(\ref{eq:paramh}).
 
We consider a service that can be given at a continuum of different
levels of quality, e.g., the acceptable delay in delivering a packet
over the Internet.
In this paper, quality is assumed to be one-dimensional (and totally
ordered). This is a severe assumption. Quality of service over the
Internet is generally considered to be described by a three-dimensional
vector: delay, latency and jitter.
In this work the quality of service is characterized by
a real number \mbox{$q \geq 1$}. A service quality of $1$ is the lowest
quality available.
How does a service of quality $q$ compare with a service of quality $1$?
The following will serve as a quantitative definition of the quality $q$ 
of a service: 
\begin{quotation}
{\bf Quality} The cost of providing a service of quality $q$
is the cost of providing $q$ services of quality $1$.
\end{quotation}
We assume significant economies of scale: the cost of providing
$w$ services of quality $1$ grows like \mbox{$w^{s}$} for some $s$,
\mbox{$0 < s < 1$}: 
the smaller the value of $s$, the larger the economies of scale.
In~\cite{FishOdly:98} one may find a discussion
of which $s$ best fits the Internet data: values between $1/2$ and $3/4$
seem reasonable.
The cost of providing $w$ services of quality $1$ is therefore:
\[
{cost} = c \: {w}^{s}
\]
where $c$ is some positive value that characterizes 
the price of the equipment. The number $c$ represents
the level of the technology, it is a {\em technological constant}.
Many studies indicate that the price of computer and Internet equipment is
decreasing at a rapid pace, probably exponentially. The dynamics of our
model is described by a decrease in the value of $c$.

The cost of providing $f(q)$ services of quality $q$ for every $q$
is:
\[
{cost} = c \: \left( \int_{1}^{\infty} \! q \: f(q) \: dq \right) ^ s.
\]
As is customary, we include the profits in the costs and will
describe the equilibrium by an equation equating costs and revenues.
Note that the equation above expresses the fact that economies of scale
are obtained only among services of the same quality and do not obtain
for performing services of two different qualities. 

The second half of our model consists of a demand curve describing
the demand for services of quality $q$ at price $p$.
We assume that the demand for such services is described by a density
function \mbox{$d(q,p)$}.
The demand for services of quality
between $a$ and $b > a$ is:
\[
{demand} = \int_{a}^{b} \! d(q,p) \: dq
\]
if the price of any such service is $p$.
Notice that we assume that services of different quality are priced
identically.
It is very difficult to come up with justifiable assumptions about
the form of the demand function \mbox{$d(q,p)$}.
One may assume that, for any fixed $q$, the value of $d(q,p)$ is
decreasing in $p$, and even approaches zero when $p$ tends to infinity.
The rate of this decrease is much less clear.
For a fixed $p$, how should $d(q,p)$ vary with $q$?
Equivalently, what is the distribution of the demand over different qualities 
of service?
This is not clear at all. Both cases of $d(q,p)$ increasing and decreasing
in $q$ will be considered.
Fortunately, our results are very general and need not assume much about
those questions.
There are two assumptions we must make, though. They will be described
and discussed now.
The first one is that the effect of price on the demand is similar at
all quality levels, or that the distribution of demand over $q$ is the
same at all prices. 
We assume that the function $d$ is the product
of two functions, one that depends only on $q$ and one that depends only
on $p$. In the absence of information (difficult to obtain) on the
exact form of the demand curve, this sounds like a very reasonable
first order approximation.
\begin{equation}
\label{eq:decoup}
{\bf Decoupling} \ \ d(q,p) = f(q) \: h(p)
\end{equation}
The function $f$ describes the distribution of the demand over different
qualities of service.
The (decreasing) function $h$ describes the effect of price on the demand.
Our results depend on this Decoupling Assumption but should
be stable under small deviations from this approximation.
The problem of studying systems in which this assumption does not hold 
is left for future work.
Our second assumption is that the function $f$ is constant in time, i.e.,
does not depend on the technological constant $c$.
Technological progress modifies demand only through price.
Our results about the dynamics of the model rely on this assumption.

To be able to prove mathematically precise results, we shall, when needed, 
assume that:
\begin{equation}
\label{eq:paramf}
f(q) = q^{\alpha}
\end{equation}
for some $\alpha$:
the larger the value of $\alpha$, the larger the relative size of the demand 
for high-quality services.
Is it clear whether $\alpha$ should be
positive or negative? Suppose e-mail packets and streaming-video packets
can be sent at the same price, would the demand for streaming-video packets
be larger than that for e-mail? Probably. It seems that a positive
value for $\alpha$ is more realistic, but
we do {\em not} make any assumption on the sign of $\alpha$.
It may be the case that a more realistic function $f$ should, first, 
for $q$'s close to one, increase, and then decrease for $q$'s above the
quality required for the most demanding applications at hand.
We shall not try to discuss this case here, but the conclusions of this paper
offer many qualitative answers even for such a case. 

For $h$, we shall always assume that
\mbox{$h(p) \geq 0$} for every \mbox{$p \geq 0$}, that $h$ is decreasing
when $p$ increases and that, when $p$ approaches $+ \infty$, 
$h(p)$ approaches zero.
Since $h(0)$ characterizes the demand at price $0$, we shall also assume,
for proper scaling, that \mbox{$h(0) = 1$}.
The following are examples of possible forms for $h$:
\begin{equation}
\label{eq:paramh}
h(p) = \frac{1}{1 + {(a \: p)}^{\beta}}, {\rm \ for \ some \ } \beta > 0, \ 
a > 0
\end{equation}
In~(\ref{eq:paramh}), $\beta$ describes the asymptotic rate of decrease 
of the demand when price increases, but $a$ describes the transient behavior of
the demand:
the larger $\beta$, the more sensitive the demand is to price.
For $\beta$ close to zero, the demand is little influenced by price, i.e.,
the demand is incompressible.
One may also consider the following:
\begin{equation}
\label{eq:exp}
h(p) = e^{-p},
\end{equation}
\begin{equation}
\label{eq:expsq}
h(p) = e^{-p^2}.
\end{equation}
As explained at the start of this Section, precise mathematical
results will assume that the functions $f$ and $h$ are defined 
by~(\ref{eq:paramf}) and~(\ref{eq:paramh}), but the qualitative results
hold for a much larger family of functions. 
\section{Beginnings}
\label{sec:begin}
\subsection{Equilibrium Equation}
\label{subsec:equilibrium}
As explained in Section~\ref{sec:intro}, we assume that the service provider
cannot provide differentiated services: it has to provide
the same service and to charge the same price for every service it accepts
to perform.
Let us, first, assume that there is no service available and that 
a service provider considers whether or not to enter the market.
In a sense, the provider is a monopolist for now.
But we shall show below that, because of possible competition, it cannot
maximize its profit.
It is clear that a provider will choose a quality
of service $b$ and a price $p$, and offer services of quality $b$
at price $p$ to any potential customer.
The potential customers are all those customers who request services of quality
\mbox{$q \leq b$}. 
Customers requesting quality above $b$ will not be served. 
If there are no competitors, one can hope to attract
the totality of the demand for services of quality \mbox{$q \leq b$}
at price $p$.
The size of the demand for service of quality $q$ (\mbox{$q \leq b$})
is \mbox{$f(q) h(p)$}. Since we provide a service of uniform quality
$b$, the equivalent number of services of quality $1$ that we shall
perform is:
\begin{equation}
\label{eq:X}
\int_{1}^{b} \! b \: f(q) \: h(p) \: dq = b \: h(p) \: A(1,b),
\end{equation} 
where we define:
\[
A(a,a') = \int_{a}^{a'} \! f(q) \: dq.
\]
Note that $A(q,q')$ is the demand for services of quality between $q$
and $q'$ if the price for those services is zero.
The cost of providing the services of~(\ref{eq:X}) is:
\begin{equation}
\label{eq:cost}
{cost} = c \: {b}^{s} \: h^{s}(p) \: A^{s}(1,b).
\end{equation}
The total demand for services is:
\begin{equation}
\label{eq:demand}
{demand} = \int_{1}^{b} \! f(q) \: h(p) \: dq = h(p) \: A(1,b).
\end{equation}
Since every service pays the same price $p$, the revenue is:
\begin{equation}
\label{eq:revenue}
{revenue} = p \: h(p) \: A(1,b).
\end{equation}
Equilibrium is characterized by:
\[
p \: h(p) \: A(1,b) = c \: {b}^{s} \: h^{s}(p) \: A^{s}(1,b),
\]
or equivalently by:
\begin{equation}
\label{eq:eq}
p \: h^{1 - s}(p) = c \: {b}^{s} \: A^{-(1-s)}(1,b).
\end{equation}
Equation~(\ref{eq:eq}) characterizes the price $p$ as a function of the quality
of service $b$. Most of the remainder of this paper is devoted to the study
of this equation.

\subsection{Existence of an equilibrium}
\label{subsec:price}
\begin{figure}[htbp]
\centering
\ifthenelse{\boolean{color}}
{\includegraphics[height=8cm]{lhs2col.ps}}
{\includegraphics[height=8cm]{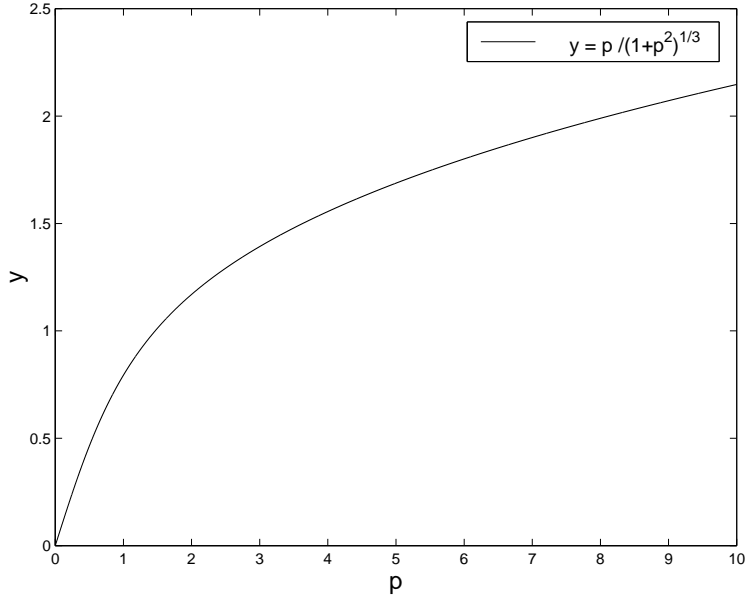}}
\caption{Left hand side of~(\ref{eq:eq}), $s = 2 / 3$, $\beta = 2$}
\label{fig:lhs2}
\end{figure}

Given $b$ and $c$, we want to solve~(\ref{eq:eq}) in the variable
$p$.
The solutions of~(\ref{eq:eq}) are obtained as the intersections
of the curve representing the left hand side and a horizontal line
representing the right hand side.
The left hand side is equal to zero for \mbox{$p=0$} and is always 
non-negative.
It is reasonable to assume that, considered as
a function of $p$, it has one of the two
following forms.
\begin{figure}[htbp]
\centering
\ifthenelse{\boolean{color}}
{\includegraphics[height=8cm]{lhs1col.ps}}
{\includegraphics[height=8cm]{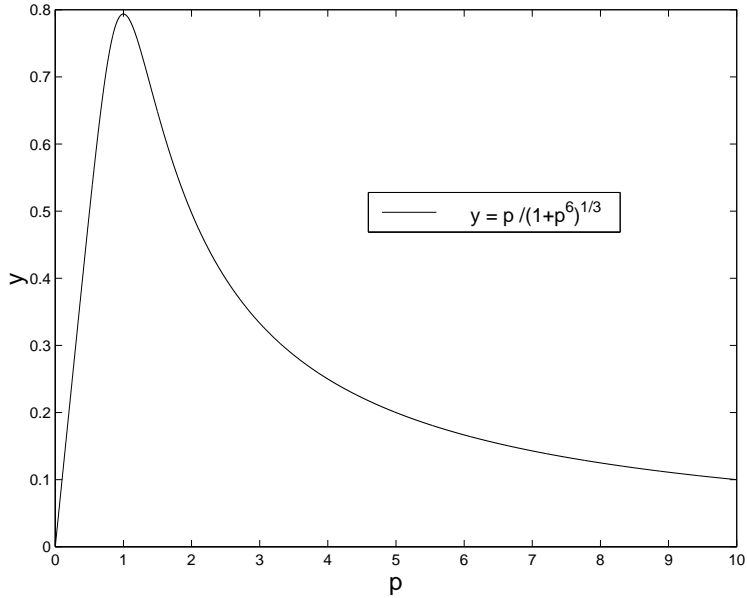}}
\caption{Left hand side of~(\ref{eq:eq}), $s = 2 / 3$, $\beta = 6$}
\label{fig:lhs1}
\end{figure}
\begin{itemize}
\item Sensitive case: \mbox{$p h^{1-s}(p)$} increases, has a maximum and 
then decreases. This seems to be the most typical case: the price has
a strong influence on the demand and therefore $h$ is such that 
\mbox{$h^{1 - s}(p)$} decreases more rapidly than $1 / p$ when $p$ is large.
This is the case, for any $s$, 
when $h$ has the form described in~(\ref{eq:exp}) 
or~(\ref{eq:expsq}). If $h$ follows~(\ref{eq:paramh}), it is the case
iff \mbox{$\beta > \frac{1}{1-s}$}.
Fig~\ref{fig:lhs1} presents a typical sensitive case: 
\mbox{$s = 2 / 3$}, \mbox{$\beta = 6$} and \mbox{$a = 1$}, i.e.,
\mbox{$h(p) = \frac{1}{1 + p^{6}}$}.
\item Insensitive case: \mbox{$p h^{1-s}(p)$} 
increases always and tends to 
\mbox{$+ \infty$} when $p$ tends to \mbox{$+ \infty$}. 
This is the case if the demand is relatively insensitive to the price.
It happens if $h$ follows~(\ref{eq:paramh}) 
and \mbox{$\beta < \frac{1}{1-s}$}.
The case in which \mbox{$p h^{1-s}(p)$} increases and approaches a finite
value when $p$ tends to \mbox{$+ \infty$} is possible 
(\mbox{$\beta = \frac{1}{1-s}$}) but not generic.
Fig~\ref{fig:lhs2} presents a typical insensitive case:
\mbox{$s = 2 / 3$}, \mbox{$\beta = 2$} and \mbox{$a = 1$}, i.e.,
\mbox{$h(p) = \frac{1}{1 + p^{2}}$}.
\end{itemize}
Equation~(\ref{eq:eq}) behaves quite differently in those two cases.
Whereas in the insensitive case, (\ref{eq:eq}) has a unique solution 
whatever the values of $c$ and $b$, in the sensitive case 
we must distinguish between two situations.
\begin{itemize}
\item If the horizontal line is high, e.g., if the technological constant $c$
is large, the horizontal line does not intersect the curve and
there is no price $p$ to solve~(\ref{eq:eq}). In this case
providing a service of quality $b$ is unprofitable, even in the absence
of competition and no such service will be provided.
\item If the horizontal line representing the right hand side intersects
the curve, for each intersection there is a price 
$p$ solving~(\ref{eq:eq}). In this case,
the provider may consider providing a service of quality $b$ at any such
price $p$, but before any more sophisticated analysis, it is clear that
the price at which service of quality $b$ will be proposed (if proposed)
is the smallest of the $p$'s solving~(\ref{eq:eq}).
If a provider proposes another solution of the equation, a competitor
will come in and propose a service of the same quality at a lower price. 
\end{itemize}
In other terms: if~(\ref{eq:eq}) has no solution, no service of
quality $b$ can be profitably provided, and if~(\ref{eq:eq}) has
at least a solution then the service may be provided at a price that is
the smallest solution of~(\ref{eq:eq}).

\subsection{Quality of the service provided in a competitive environment}
\label{subsec:service}
\begin{figure}[htbp]
\begin{tabular}{cc}
\ifthenelse{\boolean{color}}
{\includegraphics[height=6cm]{rhs1col.ps} &
\includegraphics[height=6cm]{rhs2col.ps}}
{\includegraphics[height=6cm]{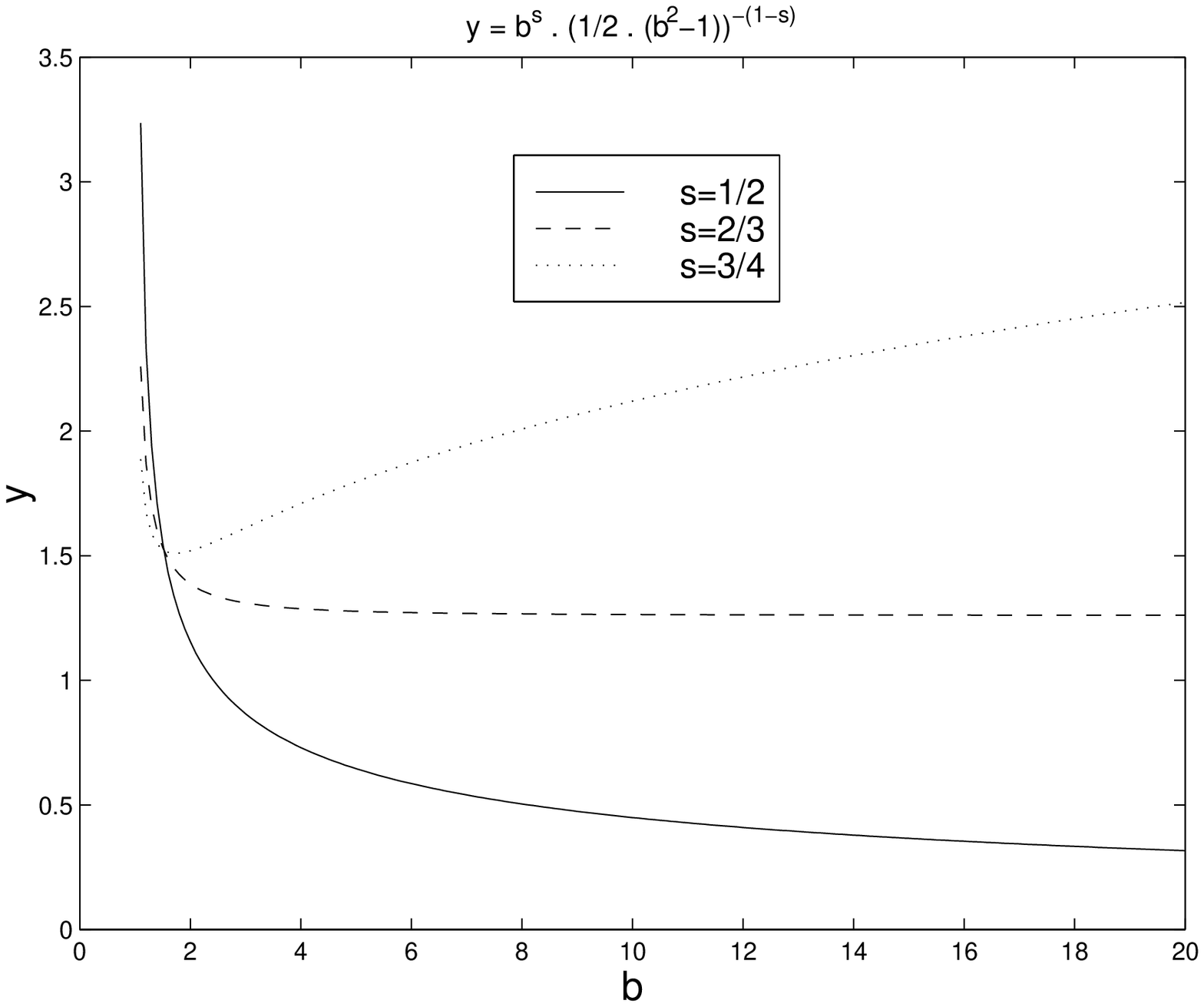} &
\includegraphics[height=6cm]{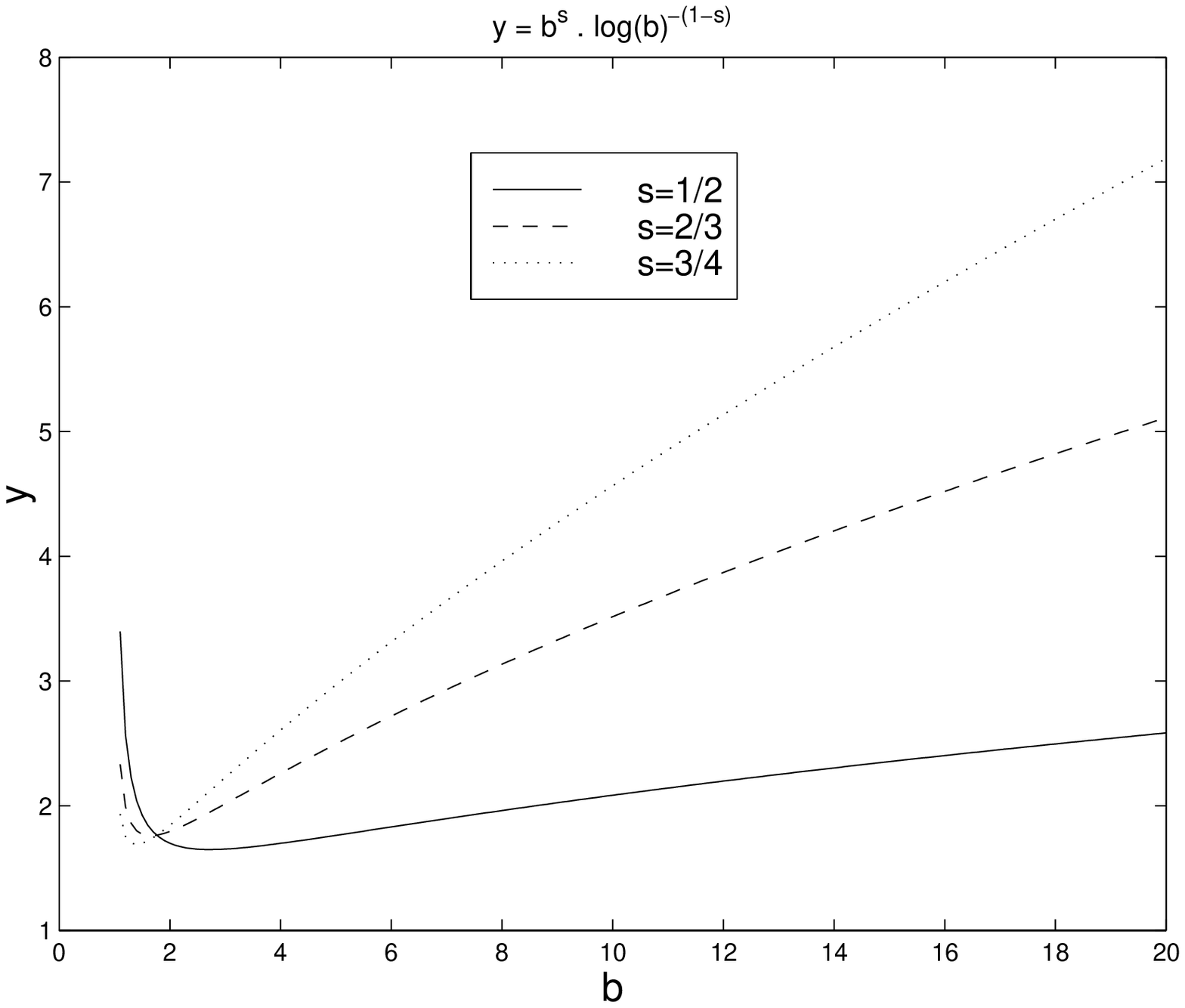}}
\end{tabular}
\caption{Right hand side of~(\ref{eq:eq})}
\label{fig:rhs}
\end{figure}

Suppose a service provider finds it is in a situation 
in which~(\ref{eq:eq}) has solutions for some values of $b$.
It may consider providing a service of quality $q$ for any one of those
values. The price it will charge, the traffic and the revenue obtained 
depend on the quality $q$ chosen. To understand the choice that
a provider has to make, we need to study~(\ref{eq:eq})
in some more depth.
Assume that $c$ is given. One would like to know which qualities
of service can be profitably provided, i.e.,
one wants to consider the set of 
$b$'s for which the Equation has a solution.
The left hand side of the Equation has been studied in 
Section~\ref{subsec:price}: let us study the right hand side.
Under very general circumstances, for \mbox{$b \rightarrow {1}^{+}$}, 
\mbox{$A^{-(1-s)}(1,b)$} approaches $+ \infty$.
Figure~\ref{fig:rhs} describes the behavior of the
right hand side of~(\ref{eq:eq}) as a function of $b$ for different
choices of $f$ (i.e. $\alpha$) and $s$. 
The number $c$ is chosen to be $1$ in all cases.
In the leftmost graph \mbox{$\alpha = 1$} and in the rightmost graph
\mbox{$\alpha = -1$}.
The function \mbox{$b^{s} A^{-(1-s)}(1,b)$} also approaches $+ \infty$
for $b$ approaching $1$ and, essentially, it may present any one of 
two possible behaviors: either it decreases always, or it decreases
until it gets to a minimum and then increases without bounds.
The case it decreases to a minimum and then increases and approaches
a finite value when $b$ tends to \mbox{$+ \infty$} is possible but not
generic.
These two possible behaviors for the function 
\mbox{${b}^{s} \: A^{-(1-s)}(1,b)$} determine two very different regimes.
\begin{definition}
\label{def:regime}
In the first regime, the Universal Class regime (UC regime), 
the right hand side of~(\ref{eq:eq})
is a decreasing function of $b$. In this regime the set of $b$'s 
for which~(\ref{eq:eq}) has a solution is an interval of the form 
\mbox{$[q_{1} , + \infty[$}, for some \mbox{$q_{1} > 1$}. 
Any service of quality at least $q_{1}$ may be profitably provided.
In the second regime, the Differentiated Classes regime (DC regime), 
the right hand
side of~(\ref{eq:eq}) decreases, reaches a minimum and increases.
One must then distinguish between the sensitive and the insensitive cases.
In the sensitive case, the set of $b$'s for which~(\ref{eq:eq}) 
has a solution is either empty or an interval of the form 
\mbox{$[q_{1} , q_{2}]$}, for some \mbox{$q_{2} \geq q_{1} > 1$}.
This regime (DC, sensitive case) will be called the BDC (bounded DC)
sub-regime.
In the insensitive case, (\ref{eq:eq}) has a solution for any $b$
in the interval \mbox{$[1 , + \infty [$}.
This regime (DC, insensitive case) will be called the UDC (unbounded DC)
sub-regime.
\end{definition}
Suppose a service provider finds it is in a situation in which~(\ref{eq:eq}) 
has solutions for some interval of qualities. 
We know it should choose a quality $q$ in this interval,
offer a service of quality $q$ and propose this service at the price $p$
that is the smallest solution to~(\ref{eq:eq}).
Which $q$ should our provider choose?
To each choice for $q$ corresponds a price $p_{q}$.
The number of services performed is given by 
\[
{demand} = h(p_{q}) \: A(1,q).
\]
Since every service performed is a service of quality $q$,
and every such service is equivalent to $q$ services of quality $1$,
the weighted traffic is:
\[
{traffic} = q \: h(p_{q}) \: A(1,q).
\]
The revenue is:
\[
{revenue} = p_{q} \: h(p_{q}) \: A(1,q).
\]

\begin{figure}[htbp]
\centering
\ifthenelse{\boolean{color}}
{\includegraphics[height=8cm]{ptraffrevdccol.ps}}
{\includegraphics[height=8cm]{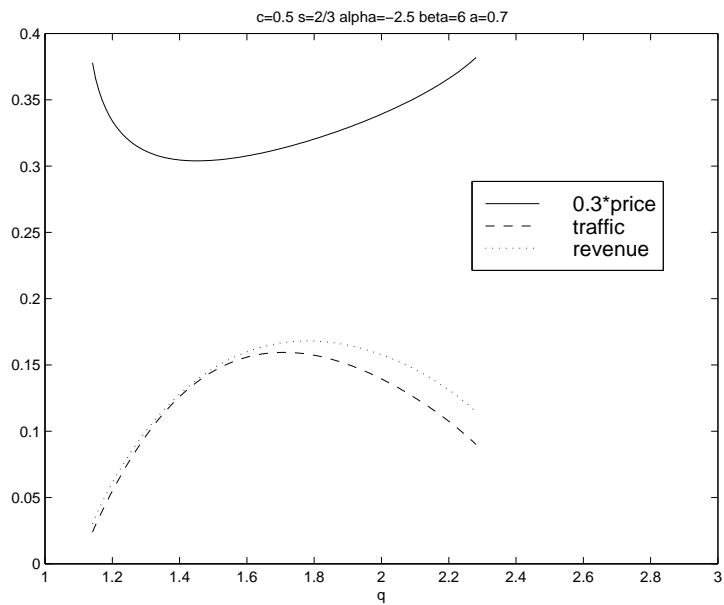}}
\caption{Price, Traffic, Revenue as functions of $q$ in the DC regime.
Service can be provided only for $q$ in a bounded interval}
\label{fig:ptraffrevDC}
\end{figure}

Consider, first, a case typical of the DC regime. 
In Figure~\ref{fig:ptraffrevDC} the price $p_{q}$, the traffic 
and the revenue are described as functions 
of the quality $q$ of the service chosen by the provider.
The service can be provided only for a bounded interval of qualities.
Notice that, in this interval, the price decreases rapidly 
and then increases slowly,
that traffic and revenue increase and then decrease, and that the
minimum price is obtained for a quality inferior to the quality
that maximizes the revenue.

\begin{figure}[htbp]
\centering
\ifthenelse{\boolean{color}}
{\includegraphics[height=8cm]{ptraffrevuccol.ps}}
{\includegraphics[height=8cm]{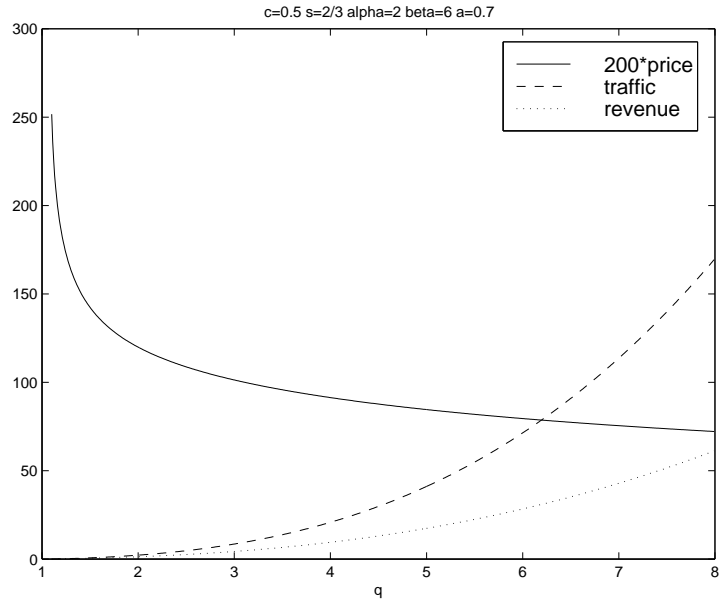}}
\caption{Price, Traffic, Revenue as functions of $q$ in the UC regime}
\label{fig:ptraffrevUC}
\end{figure}

A case typical of the UC regime is described in
Figure~\ref{fig:ptraffrevUC}.
By our discussion of~(\ref{eq:eq}), it is clear that $p_{q}$ is
a decreasing function of $q$. Notice that traffic and revenue increase 
with $q$.

What is the quality of service most profitable?
Since the revenue includes the profits, one may assume, at first sight,
that the first provider to propose such a service
will choose the quality $q$ that maximizes its revenue.
This analysis is clearly incorrect in a competitive situation.
If a provider decides to offer the service that maximizes its revenue,
a competitor provider may come in and offer a service of lower quality 
at a lower price and steal the demand for the lower quality services.
Therefore, the price at which the service will be offered is the
minimum price possible: the minimum of the function $p_{q}$ on its interval of
definition.
\begin{claim}
The quality $q$ of the service provided is the $q$ for which $p_{q}$
is minimal.
\end{claim}

This remark turns out to be of fundamental importance because
this quality $q$, contrary to the quality that maximizes revenue,
does not depend on the technological constant $c$ or on the function $h$
that describes the dependence of the demand on the price charged. 
It depends only on the function $f$ that describes the distribution
of the demand over the different qualities of service and on the size
of the economies of scale described by the number $s$.
Indeed, this quality $q$ is the value of $b$ that minimizes
\mbox{$b^{s} A^{-(1-s)}(1,b)$} in the DC regime and it is $+ \infty$ in
the UC regime.
There is, therefore, a sharp distinction between the two regimes:
in the DC regime a service of finite quality is provided,
whereas, in the UC regime, the quality of service provided
is high enough to please every customer, i.e., unbounded.
\begin{law}[First Law]
Whether or not some service is provided does not depend on the function
$h$ that describes the way demand depends on price.
It depends on the technological constant $c$, the size of the economies
of scale $s$, and on the function $f$ that describes the distribution of 
demand over different qualities of service.
If some service is provided its quality depends only on $s$ and $f$, 
it does not depend on $c$ or $h$.
In the UC regime, a single class serves requests for services of arbitrary
quality and it is given for free.
In the DC regime, the service of lowest quality has a finite quality
and does not cater for high-end customers.
The prevailing regime depends on the form of the economies
of scale and on the form of the distribution of the demand over different
qualities, it does not depend on the price of the equipment or on the
way prices influence demand.
\end{law}
In Section~\ref{sec:border}, we shall, first, mark the boundary between
these two regimes. Then, both regimes: UC and DC will be studied in turn.

\section{Between the two regimes}
\label{sec:border}
In this Section, we shall assume that $f$ has the form described 
in~(\ref{eq:paramf}), for some $\alpha$, and study the exact boundaries
of the UC and DC regimes.
The regime is determined by the behavior of the function:
\[
w(b) = b^{s} \: \left( \int_{1}^{b} \frac {dq}{q^{\alpha}} \right)^{-(1-s)},
\]
for \mbox{$b \geq 1$}.
If $w(b)$ has a minimum, then the regime is DC, if it always decreases,
the regime is UC.
Three cases must be distinguished.
\[
{\rm If \ } \alpha < -1 , \ \ 
w(b) = \frac {b^{s}} {{ \left( -\alpha - 1 \right) }^{-(1-s)}} \;
{ \left( 1 - b^{\alpha+1} \right) }^{-(1-s)}.
\]
\[
{\rm If \ } \alpha =  -1 , \ \ 
w(b) = b^{s} \: \ln^{-(1-s)} b,
\]
and 
\[
{\rm if \ } \alpha > -1 , \ \ 
w(b) = \frac {b^{s}} {{ \left( \alpha + 1 \right) }^{-(1-s)}} \;
{ \left( b^{\alpha+1} - 1 \right) }^{-(1-s)}.
\]
For \mbox{$\alpha < -1$}, we have:
\[
w'(b) = \frac {b^{-(1-s)} \: {\left( 1 - b^{\alpha+1} \right)}^{-(2-s)} }
{{ \left( -\alpha - 1 \right) }^{-(1-s)}}
\; \left[ s (1 - b^{\alpha+1}) - (1-s) b (-\alpha-1) b^{\alpha} \right] ,
\]
\[
w'(b) = \frac {b^{-(1-s)} \: {\left( 1 - b^{\alpha+1} \right)}^{-(2-s)} }
{{ \left( -\alpha - 1 \right) }^{-(1-s)}}
\; \left[ (-(2+\alpha) s + \alpha + 1) \: b^{\alpha+1} + s \right] ,
\] 
which approaches $s$ and is therefore positive for large $b$'s.
We conclude that for any \mbox{$\alpha < -1$} the regime is DC.

For \mbox{$\alpha = -1$}, 
\[
w'(b) = b^{s-1} \: \ln^{-(2-s)} b \: 
\left[ s \ln b - (1-s) \right],
\]
which is also positive for large $b$'s.
For \mbox{$\alpha = -1$} the regime is DC.

\begin{figure}[htp]
\centering
\ifthenelse{\boolean{color}}
{\includegraphics[height=8cm]{ucdccol.ps}}
{\includegraphics[height=8cm]{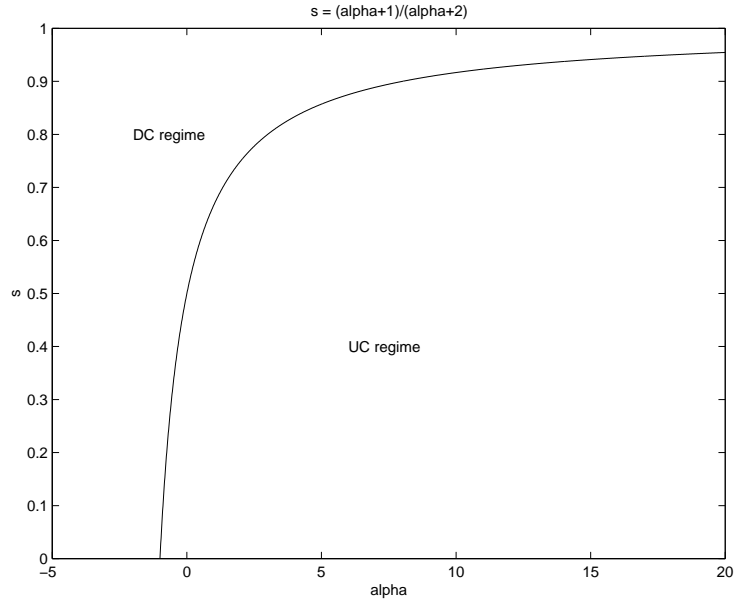}}
\caption{Boundary between the UC and DC regimes}
\label{fig:UCDC}
\end{figure}

For \mbox{$\alpha > -1$}, we have:
\[
w'(b) = \frac {b^{-(1-s)} \: {\left( b^{\alpha+1} - 1 \right)}^{-(2-s)} }
{{ \left( \alpha + 1 \right) }^{-(1-s)}}
\; \left[ s (b^{\alpha+1}-1) - (1-s) b (\alpha+1) b^{\alpha} \right]
\]
\[
w'(b) = \frac {b^{-(1-s)} \: {\left( b^{\alpha+1} - 1 \right)}^{-(2-s)} }
{{ \left( \alpha + 1 \right) }^{-(1-s)}}
\; \left[ ( (2+\alpha) s - \alpha - 1 ) \: b^{\alpha+1} - s \right]
\]
For large $b$'s, $w'(b)$ has the sign of \mbox{$(2+\alpha) s - \alpha - 1$}.
We may now conclude our study.
\begin{claim}
\label{claim:DCUC}
\[
{\rm the \ regime \ is \ DC \ iff \ } \alpha \leq -1 {\rm \ or \ }
s > \frac{\alpha+1}{\alpha+2},
\]
\[
{\rm the \ regime \ is \ UC \ iff \ } \alpha > -1 {\rm \ and \ }
s \leq \frac{\alpha+1}{\alpha+2},
\]
\end{claim}
The boundary between the two regimes can be seen in Figure~\ref{fig:UCDC}. 
Informally, if the economies of scale are very substantial (small $s$)
and if high quality services are in big demand ($f$ grows rapidly),
the UC regime prevails.
Otherwise, there is an upper bound to the quality of profitable services. 
There is no profitable way to provide
for a service of arbitrary quality because its cost would
deter too many potential customers.
\section{The universal class regime}
\label{sec:univclass}
\subsection{A paradoxical regime}
\label{subsec:paradox}
Let us study the UC regime. As discussed in Section~\ref{subsec:service},
Figure~\ref{fig:ptraffrevUC} and in our first Law, in this regime,
the right hand side of~(\ref{eq:eq}) is as close to zero
as one wishes if one takes $b$ to be large enough.
For any value of the technological constant $c$, even a very high one,
it is profitable to provide a service of infinite quality at zero price,
and therefore, for any value of the technological constant $c$, 
a unique class of service of unbounded quality is offered for free.
This seems quite a paradoxical situation. 
The demand (\ref{eq:demand}) and the cost of providing the service 
(\ref{eq:cost}) are both infinite.
The consideration of the revenue is interesting.
The revenue (see (\ref{eq:revenue}) and~(\ref{eq:eq})) is infinite,
but depends linearly on the technological constant $c$.
The revenue drops linearly with the price of the equipment.

This regime seems to fit the fat dumb pipe model of the Internet: 
there are enough
resources to give every request a treatment of the highest quality,
there is no need to give differentiated services and the price is very low
since the demand is very high and the economies of scale very substantive.

The dynamics of the UC regime under technological progress, i.e., a 
drop in the constant $c$, is quite paradoxical.
Such a drop does not significantly affect the, already negligible,
price of the service, but it affects adversely 
the (infinite) revenues of the provider.

The First Law implies that the boundary between the UC and DC regimes 
does not depend on the technological constant $c$,
therefore the answer to the question whether the fat dumb pipe model
prevails or whether a number of classes of service will develop
depends only on the relation between the economies of scale and
the distribution of demand over different qualities, it does not depend
on the price of the equipment. 
Both regimes are stable under technological progress.
\subsection{A more realistic model}
\label{subsec:realistic}
\begin{figure}[htbp]
\centering
\ifthenelse{\boolean{color}}
{\includegraphics[height=8cm]{ucrealcol.ps}}
{\includegraphics[height=8cm]{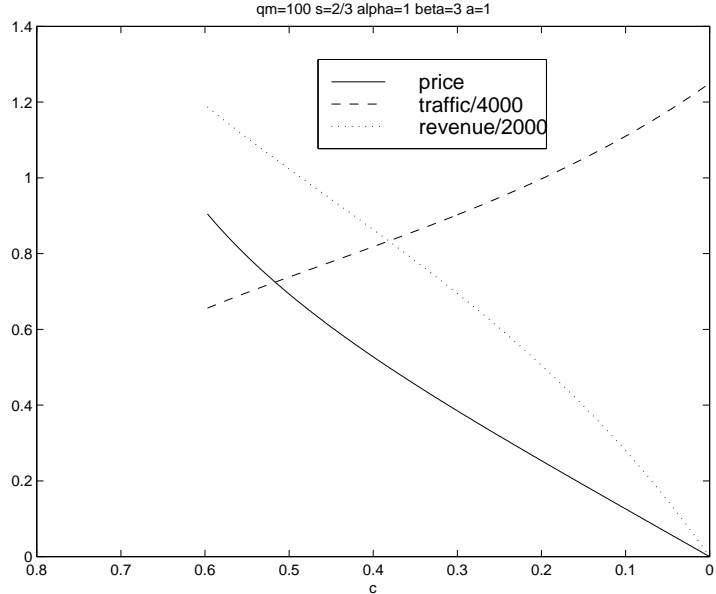}}
\caption{Evolution of price, traffic and revenue in the UC regime}
\label{fig:UCreal}
\end{figure}

The paradoxical aspects of the UC regime as described
above stem from the consideration of a demand curve that assumes 
significant demand for services of unbounded quality.
It seems more realistic to assume an upper bound to the quality requested
from services. A specific example will be given now, to explain what 
the UC regime may look like in practice.
Assume that there is an upper bound $q_{m}$ to the quality of service
requested and that the distribution of demand $f$ over different qualities 
is described by: 
\begin{equation}
\label{eq:piecewise}
f(q) = \left\{ \begin{array}{ll}
			q & \mbox{if $1 \leq q \leq q_{m}$} \\
			0	& \mbox{if $q_{m} \leq q$}
		\end{array}
	\right. 
\end{equation}
Assume also that \mbox{$s = \frac{2}{3}$} and that 
\mbox{$h(p) = \frac{1}{1+p^3}$}.
It follows from our discussion above that the service provided, if provided,
will be of quality $q_{m}$.
Therefore the equilibrium equation is:
\begin{equation}
\label{eq:speceq}
\frac{p}{(1+p^3)^{1/3}} = 2^{1/3} \: c \: {q_{m}}^{2/3} \: 
{\left( {q_{m}}^2 - 1 \right)}^{-1/3}.
\end{equation}
The left hand side of~(\ref{eq:speceq}) has a maximum value of $1$.
For any $c$ such that 
\[
c \geq 2^{-1/3} \: {q_{m}}^{-2/3} \: 
{\left( {q_{m}}^2 - 1 \right)}^{1/3}
\]
there is no profitable service.
For any $c$ smaller than this value, 
a provider will choose to provide a service of quality $q_{m}$.
The price, traffic and revenue are functions of $c$ and described in
Figure~\ref{fig:UCreal}.
As soon as some service is provided, it caters for all needs.
A unique class of service will still prevail in the presence of
a drop in the price of equipment, i.e., a decrease in $c$.
Notice that the revenue decreases when the technology constant drops.

\section{The differentiated classes regime}
\label{sec:multiple}
Let us study the DC regime now.
The best approach is to take a dynamic perspective and consider
the evolution of the system under a decrease in the technological
constant $c$.
\subsection{The BDC sub-regime}
\label{subsec:BDCregime}
\begin{figure}[htbp]
\centering
\ifthenelse{\boolean{color}}
{\includegraphics[height=8cm]{ptrbdc1col.ps}}
{\includegraphics[height=8cm]{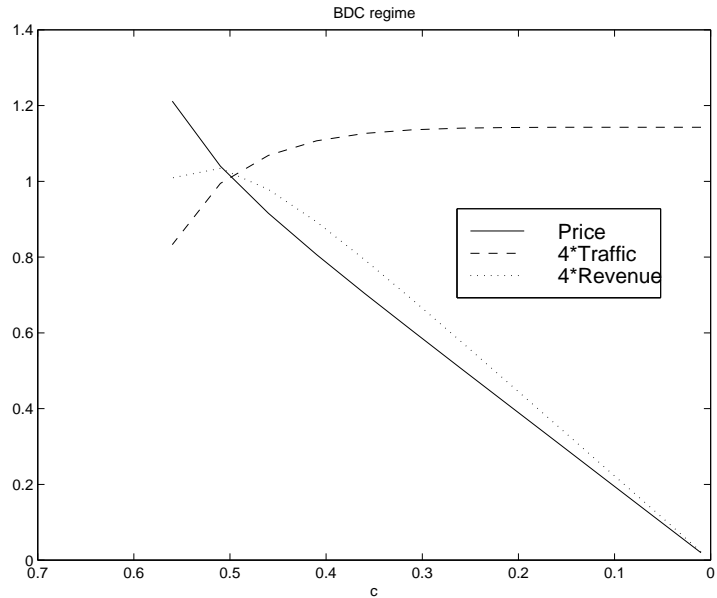}}
\caption{Evolution of price, traffic and revenue in the BDC regime, lowest class of service for $\alpha = -2.5, s = 2/3, \beta = 6$ and $a = 0.7$}
\label{fig:ptrbdc1}
\end{figure}

In the first part of this discussion we study the BDC sub-regime,
i.e., the {\em sensitive} case described in Section~\ref{subsec:price} in 
which the left hand side of~(\ref{eq:eq}) increases to a maximum 
and then decreases.
We have seen that for high $c$'s no service is provided.
When $c$ falls below a certain value, $c_{0}$,
a service of some finite quality $q_{0}$ is provided. 
From the First Law we know that, 
even in the face of further declines in $c$, the quality $q_{0}$ of the
service provided will not change.
Two main questions arise:
\begin{itemize}
\item how do the price, the traffic and the revenue of the provider of
a service of quality $q_{0}$
evolve under a further decrease in $c$?
\item will services of higher quality be offered? which? when? at what price? 
\end{itemize}
The first question is relatively easily disposed of.
Since the right hand side of~(\ref{eq:eq}) is a linear function of $c$,
the assumption that the left hand side increases to a maximum implies
that the price drops rapidly at first (for $c$ close to $c_{0}$),
and then at a slower pace. For the functions $h$ considered, 
the slope of the left hand side is one at \mbox{$p = 0$} and
the price $p$ drops essentially linearly with $c$ and approaches zero 
when $c$ approaches zero.
The demand (or the traffic) varies as \mbox{$h(p)$} and therefore
increases very rapidly at first and then approaches some finite value.
The revenue varies as \mbox{$p h(p)$} or \mbox{$c h(p)^{s}$}.
Typically it may exhibit, first, an increase, for values of $c$
at which a drop in the price of the service is over-compensated by an
increase in demand and then a decrease. For small $c$'s, the revenue
drops linearly with $c$.
Figure~\ref{fig:ptrbdc1} describes the evolution of price, traffic and revenue
for the lowest class of service, the first to be offered on the market.

The second question is deeper. Since a service of quality $q_{0}$ will
be provided at a price we have just studied, it is clear that the
only customers who could be interested in another service are those
requesting a service of quality superior to $q_{0}$.
Will such a service, of quality \mbox{$b \geq q_{0}$} be provided?
The equilibrium equation now becomes:
\begin{equation}
\label{eq:eq2}
p \: {h(p)}^{1 - s} = c \: {b}^{s} \: A(q_{0},b)^{-(1-s)}.
\end{equation}
Notice that~(\ref{eq:eq2}) is very similar to~(\ref{eq:eq}).
A service of quality $b$ may be profitably offered iff~(\ref{eq:eq2})
has a solution. See Section~\ref{subsec:price} for a discussion:
the situation is the same. There is a value $c_{1}$ above which 
there is no solution and below which there is a solution.
Since we are in the sensitive case,
the left hand side has a maximum and 
the price of the service, of quality $b$,
if provided, will be the smallest $p$ satisfying~(\ref{eq:eq2}).

The discussion of which service will be provided is similar to the
discussion found in Section~\ref{subsec:service}.
The behavior of the right hand side of~(\ref{eq:eq2})
as a function of $b$ is the same as that of~(\ref{eq:eq}).
We have assumed we are in the DC regime, and therefore there is a finite
value $q_{1}$ for $b$ that minimizes 
\mbox{${b}^{s} \: A(q_{0},b)^{-(1-s)}$}. 
The service provided will be of a quality $q_{1}$.

\begin{figure}[htbp]
\centering
\ifthenelse{\boolean{color}}
{\includegraphics[height=8cm]{ptrbdc2col.ps}}
{\includegraphics[height=8cm]{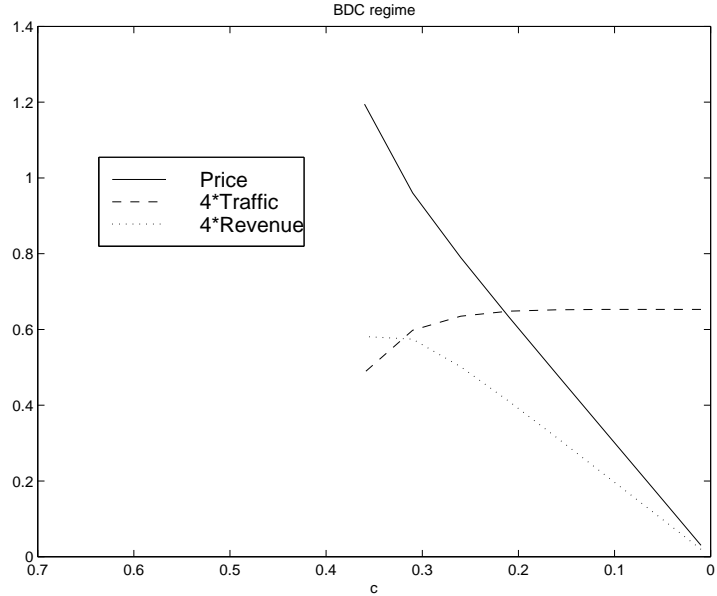}}
\caption{Evolution of price, traffic and revenue in the BDC regime, next to lowest class of service for $\alpha = -2.5, s = 2/3, \beta = 6$ and $a = 0.7$}
\label{fig:ptrbdc2}
\end{figure}

In the case $f$ has the form described in~(\ref{eq:paramf}), the
computation of $q_{1}$ is simplified by a change of variable
\mbox{$ q = q_{0} \: q'$}.
\[
A(q_{0},b) = \int_{q_{0}}^{b} \! f(q) \: dq = \int_{1}^{b / q_{0}} \! 
f(q_{0} q') \: q_{0} \: dq' = q_{0}^{\alpha+1} A(1,b/q_{0})
\]
and the equation becomes:
\[
p \: {h(p)}^{1 - s} = c' \: {(b / q_{0})}^{s} \: A(1,b / q_{0})^{-(1-s)},
{\rm \ for \ } c' = c \: q_{0}^{-(\alpha+1)(1-s)+s}.
\]
Notice that 
\[
-(\alpha+1)(1-s)+s = (\alpha+2)s - (\alpha+1) > 0
\]
by Claim~\ref{claim:DCUC} since we are in the DC regime and therefore 
\mbox{$c' > c$}.
Equation~(\ref{eq:eq2}) is the same as~(\ref{eq:eq}), for a larger
technological constant, after our change of variable.
This implies that the service provided will be of quality $q_{1}$ with
\[
q_{1} / q_{0} = q_{0} , {\rm \ i.e., \ } q_{1} = q_{0}^{2}.
\]
It will appear when
\[
c' = c_{0} , {\rm \ i.e., \ } c_{1} = c_{0} \: q_{0}^{(\alpha+1)(1-s)-s}.
\]
The general dynamic picture is now clear.
Different classes of service appear as the technological constant $c$
decreases. The $n$'th class (\mbox{$n \geq 0$}) appears when 
\mbox{$c = c_{0} \: q_{0}^{n(-s+(\alpha+1)(1-s))}$} and it is of quality
\mbox{${q_{0}}^{n}$}. For any $c>0$, only a finite number of qualities of
service are proposed and no service is provided for request of high 
quality service. Since, from Section~\ref{sec:border}, we may also
easily compute $q_{0}$ and see that:
\[
{\rm if \ } \alpha < -1 , \ \ \ \  q_{0} = 
\left( \frac{s}{(2+\alpha) s - \alpha - 1} \right) ^ {\frac{1}{\alpha+1}},
\]
\[
{\rm if \ } \alpha = -1 , \ \ \ \ q_{0} = {e}^{(1-s)/s},
\]
\[
{\rm if \ } \alpha > -1 , \   \ \ \ q_{0} = 
\left( \frac{s}{(2+\alpha) s - \alpha - 1} \right) ^ {\frac{1}{\alpha+1}},
\]
the picture is complete.
Figure~\ref{fig:ptrbdc2} describes the price, traffic and revenue in the second
class of service as functions of the technological constant $c$, for the same
constants as in Figure~\ref{fig:ptrbdc1}.
Compare with Figure~\ref{fig:ptrbdc1} and see that the qualitative evolution
is similar in both classes.
The ratio of prices in two neighboring classes
is interesting to study. 
The ratio between the prices of service in a class and the class just below,
as a function of $c$, has the same behavior and the same values 
(for different $c$'s) for all classes.
Let the classes of services are described by qualities: 
\mbox{$q_{0} < q_{1} < \ldots < q_{i} < \ldots$} and appearing 
for the values of $c$:
\mbox{$c_{0} > c_{1} > \ldots > c_{i} > \ldots$}.
Let \mbox{$p_{0}(c) < p_{1}(c) < \ldots < p_{i}(c) < \ldots$} be the prices
of the corresponding services.
The behavior of \mbox{$p_{1}(c) / p_{0}(c)$} for $c$ 
between $c_{0}$ and $c_{1}$
is the same as the behavior of  \mbox{$p_{2}(c) / p_{1}(c)$} for $c$ between 
$c_{1}$ and $c_{2}$.
The price of a newly created class (of high quality)
drops (with $c$) more rapidly than that of an older class of lower quality
and therefore the ratio \mbox{$p_{i+1}(c)/p_{i}(c)$}, 
larger than $1$, decreases with $c$.
Under the assumptions above, it approaches 
\[
\left( \frac{s}{(2+\alpha)\: s - \alpha - 1} \right)^
{\frac{(2+\alpha)\: s - \alpha - 1}{\alpha + 1}}
\]
when $c$ approaches zero.
For \mbox{$\alpha = -2$} and \mbox{$s = 1/2$}, this gives a value of $2$,
the ratio used by the Paris Metro system when it offered two classes
of service, and quite typical of other transportation systems.
\subsection{The UDC sub-regime}
\label{subsec:UDCregime}
\begin{figure}[htbp]
\centering
\ifthenelse{\boolean{color}}
{\includegraphics[height=8cm]{ptrudc1col.ps}}
{\includegraphics[height=8cm]{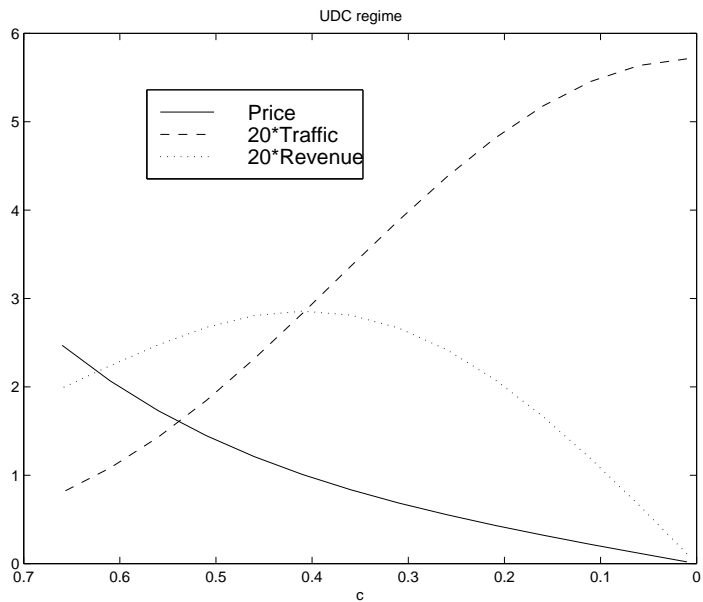}}
\caption{Evolution of price, traffic and revenue in the UDC regime, lowest class of service for $\alpha = -2.5, s = 2/3, \beta = 2$ and $a = 1$}
\label{fig:ptrudc1}
\end{figure}

\begin{figure}[htbp]
\centering
\ifthenelse{\boolean{color}}
{\includegraphics[height=8cm]{ptrudc2col.ps}}
{\includegraphics[height=8cm]{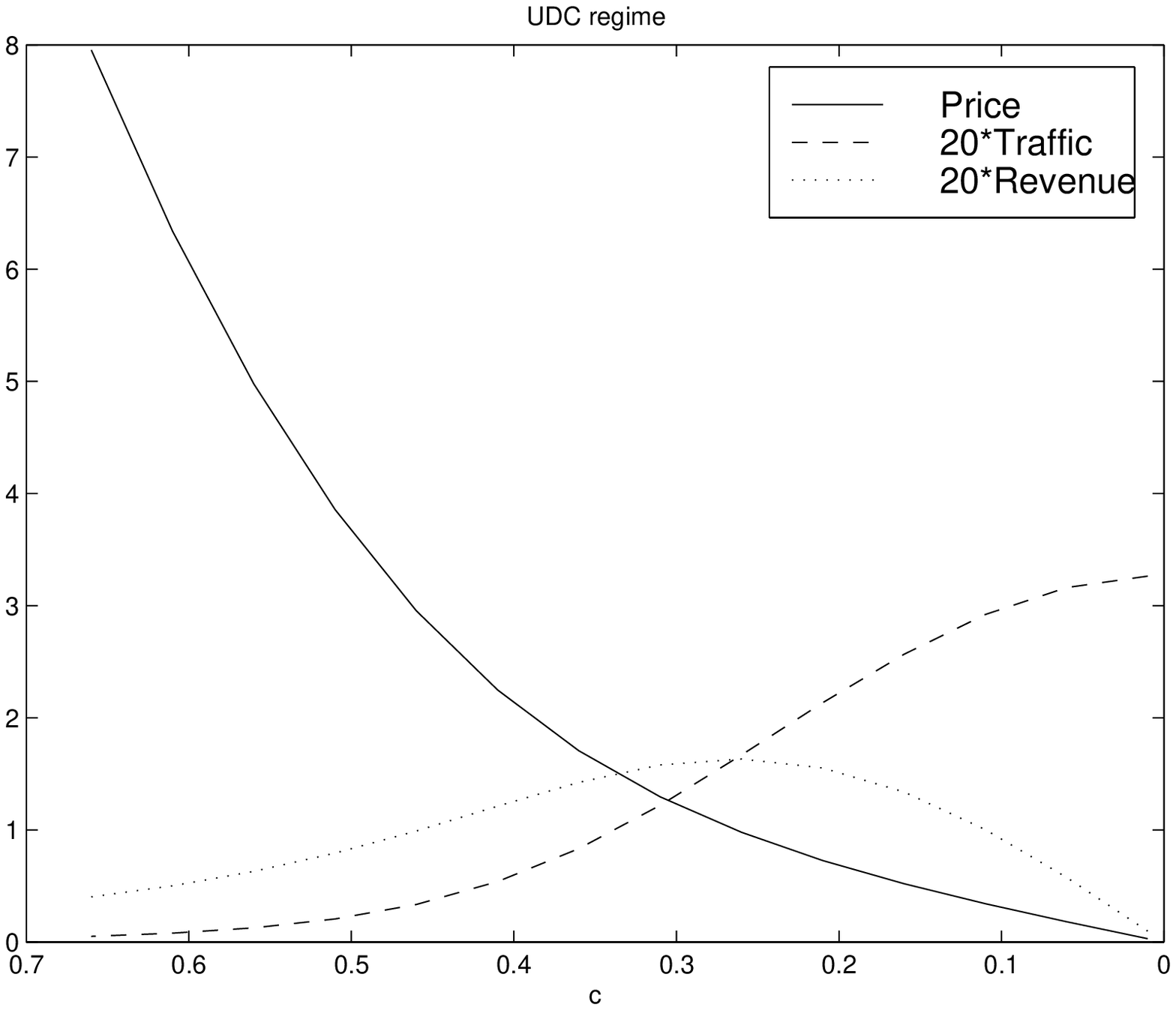}}
\caption{Evolution of price, traffic and revenue in the UDC regime, next to lowest class of service for $\alpha = -2.5, s = 2/3, \beta = 2$ and $a = 1$}
\label{fig:ptrudc2}
\end{figure}

We may now understand what happens in the insensitive case: 
the left hand side
of~(\ref{eq:eq}) grows without bound. In this case, for any $c$,
an infinite number of qualities: \mbox{$q_{0}, q_{1}, \ldots , q_{i} , \ldots$}
of service is offered. The qualities offered are constant, they do not
depend on $c$. Figures~\ref{fig:ptrudc1} and~\ref{fig:ptrudc2}
show the first two classes for a generic case.
\subsection{The general picture}
\label{subsec:genpic}
\begin{law}[Second Law]
Under a decrease in the technological constant $c$, i.e., under technological
progress and a drop in the price of equipment:
\begin{itemize}
\item In the UC regime, at any time, i.e., for any value of $c$, 
a unique class serves requests for arbitrary quality. 
\item In the DC regime,
\begin{itemize}
\item in the BDC sub-regime, no service is provided at first, 
then a service of finite quality is provided, 
later on another service of higher quality will be
proposed, and so on. For any quality $q$ a service of quality $q$ will
be proposed at some time (and on), but at any time there are qualities
of services that are not provided. 
\item in the UDC sub-regime, at any time, an infinite number of classes of service
are proposed and requests for arbitrary quality are taken care of.
\end{itemize}
\end{itemize}
\end{law}
This Second Law contradicts the folk wisdom that says that the Fat Dumb Pipe 
(FDP) model
for the Internet will collapse when prices become low enough to enable
very high quality services. It says that if the FDP model is the
one that prevails when $c$ is high, it will continue to prevail after
a decrease in $c$.
A drop in the price of equipment cannot cause the breakdown of a system
based on a universal class of service.

\section{Further work and conclusions}
\label{sec:conclusion}
The most intriguing challenges are a theoretical one and an experimental one.
The first one concerns the study of the case in which the Decoupling of 
Equation~\ref{eq:decoup}
does not hold, or holds only approximately.
The second one is the study of real systems, in particular the Internet,
and the evaluation of the parameters of interest for them.

The main conclusion of this work is that the parameters that determine
the type of the prevailing market are the size of the economies of scale
and the distribution of demand over the range of qualities.
The sensitivity of demand to price and the changes in the price of
equipment bear a lesser influence.

This paper made two severe assumptions.
It assumes perfect competition, and it assumes that economies of scale
do not aggregate over different qualities of service. Further work should
relax those assumptions.
\bibliographystyle{plain}
\bibliography{../my}

\end{document}